\documentclass[pra,twocolumn,showpacs]{revtex4}
\usepackage{graphicx,amssymb}

\begin{document}
\title{Generating Polarization-Entangled Photon Pairs with Arbitrary Joint Spectrum}

\author{Zachary~D.~Walton}
\email{walton@bu.edu} \homepage[Quantum Imaging Laboratory
homepage:~]{http://www.bu.edu/qil}

\author{Alexander~V.~Sergienko}

\author{Bahaa~E.~A.~Saleh}

\author{Malvin~C.~Teich}
\affiliation{Quantum Imaging Laboratory, Department of Electrical
\& Computer Engineering, Boston University, 8 Saint Mary's Street,
Boston, Massachusetts 02215-2421}


\begin{abstract}
We present a scheme for generating polarization-entangled photons
pairs with arbitrary joint spectrum. Specifically, we describe a
technique for spontaneous parametric down-conversion in which both
the center frequencies and the bandwidths of the down-converted
photons may be controlled by appropriate manipulation of the pump
pulse. The spectral control offered by this technique permits one
to choose the operating wavelengths for each photon of a pair
based on optimizations of other system parameters (loss in optical
fiber, photon counter performance, etc.).  The combination of
spectral control, polarization control, and lack of group-velocity
matching conditions makes this technique particularly well-suited
for a distributed quantum information processing architecture in
which integrated optical circuits are connected by spans of
optical fiber.
\end{abstract}

\maketitle

\section{INTRODUCTION}

Spontaneous parametric down-conversion (SPDC) has proven to be an
excellent technology for quantum communication, with SPDC photons
functioning as ``flying qubits.''  The discovery by Knill {\it et
al.}~\cite{Knill01} that linear optics and single photon detectors
are sufficient for scalable quantum computation has opened the
possibility that SPDC may also be useful for quantum computation.

As the proposals for quantum information processing with SPDC
become more sophisticated, the technical demands placed on SPDC
sources become more stringent.  For example, a quantum key
distribution experiment based on polarization entanglement
requires that two two-photon amplitudes ($|\mbox{HH}\rangle$ and
$|\mbox{VV}\rangle$, for example) be made indistinguishable. The
spectral properties of the two photons are only important if they
are correlated to the polarization degree of freedom.  A more
stringent form of indistinguishability is typically required for
quantum computation with linear optics: it must be impossible to
determine which source produced a certain photon after that photon
emerges from a beamsplitter. This in turn requires that all of the
photons' properties be controlled such that it is impossible to
learn any information about the identity of a given photon's
source.  Photon pairs produced by SPDC are often correlated in one
or more of their properties (frequency, direction, etc.). These
correlations can destroy the requisite indistinguishability by
enabling one to learn about one photon by performing measurements
on its twin~\cite{Uren03}.

In this paper, we describe an SPDC source that produces photon
pairs that have arbitrary correlation in frequency. The source we
propose enables an unusual flexibility in the control of the
marginal spectra of the SPDC photons. Specifically, our source can
produce frequency-uncorrelated photon pair in which the center
frequency and the bandwidth of each photon is controlled
independently, regardless of the nonlinear material's dispersion
curves. This makes the source well-suited for applications that
span quantum communication and computation, such as
teleportation~\cite{Bennett93} and entanglement
swapping~\cite{Zukowski93}. In these applications, one photon of a
pair (the ``communication'' photon) is often sent to another party
through a long span of optical fiber, while the other photon (the
``computation'' photon) is analyzed and detected in a localized
interferometer. It is desirable that the communication photon be
narrow-band (such that effects like polarization mode dispersion
are mitigated) and have a wavelength in the infrared (where
optical fiber is least lossy). Contrariwise, the computation
photon should be broad-band (such that it can be used in
interferometers with small path-length differences) and have a
wavelength suited for high-efficiency single-photon counters.

The paper is organized as follows. In section~\ref{gapmspdc}, we
introduce the new technique and demonstrate that it permits the
generation of photon pairs with arbitrary joint spectrum. In
section~\ref{example}, we show how this spectral control can be
combined with polarization entanglement by considering a specific
example involving a BBO waveguide.  In section~\ref{int}, we
discuss the possibility of using this source as part of a
distributed quantum information processor based on integrated
optics.  Finally, we summarize our results in
section~\ref{conclusions}.

\begin{figure*}
\includegraphics{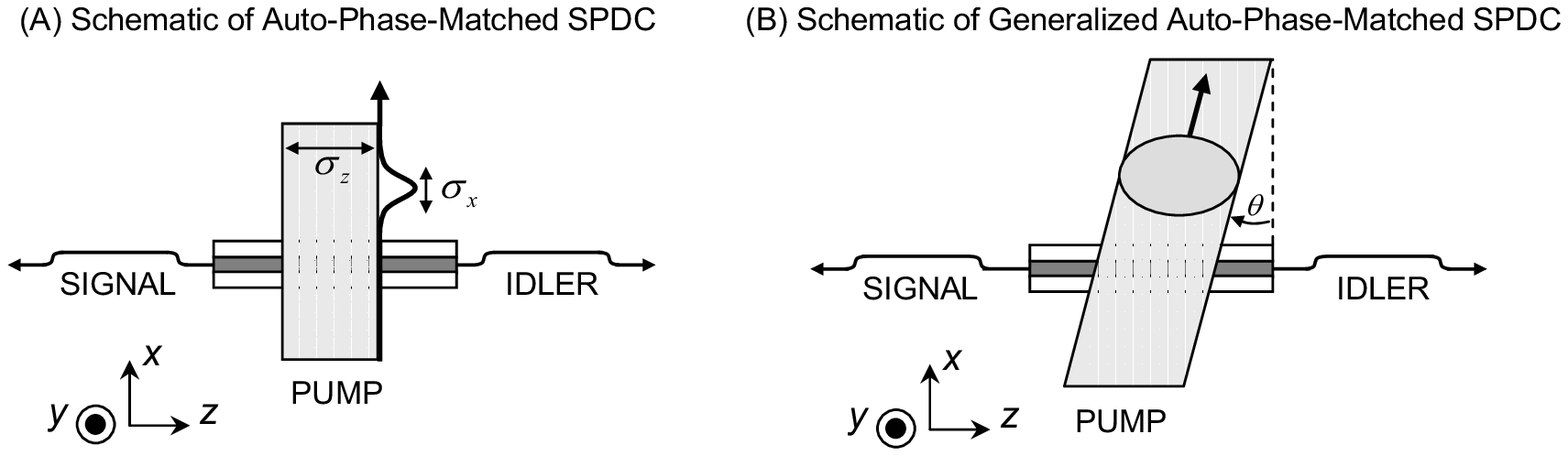}
\caption{Schematics of (A) auto-phase-matched
SPDC~\protect\cite{Walton03c} and (B) generalized
auto-phase-matched SPDC. In both schemes, a transverse pump pulse
stimulates the creation of a pair of counter-propagating photons
in a single-mode, nonlinear waveguide.   In (A), the pump pulse is
cross-spectrally pure and constrained in its direction (normal to
the waveguide).  This restricts the spectral properties of the
SPDC photons.  In (B), these constraints on the pump pulse are
relaxed. One can exploit this freedom to independently control the
center frequency and bandwidth of each SPDC photon, while
satisfying the constraint that the photons be uncorrelated in
frequency.}\label{setup}
\end{figure*}

\section{GENERALIZED AUTO-PHASE-MATCHED SPDC}\label{gapmspdc}

Our source is a generalization of the design we previously
introduced under the name {\it auto-phase-matched}
SPDC~\cite{Walton03c}; thus, we name the new scheme {\it
generalized} auto-phase-matched SPDC. Like the original scheme,
the new scheme features counter-propagating SPDC created in a
single-mode nonlinear waveguide by a transverse pump pulse. In the
original scheme (Fig.~\ref{setup}A), the pump pulse is
cross-spectrally pure ({\it i.e.} the complex envelope factors
into separate functions of space and time), and impinges on the
waveguide at normal incidence. In the new scheme
(Fig.~\ref{setup}B), the pump pulse may have cross-spectral
correlations, and may approach the waveguide at non-normal
incidence.  With these constraints on the pump pulse relaxed, the
center frequency and bandwidth of each SPDC photon may be
controlled independently.

In typical SPDC experiments, a monochromatic pump beam is used. In
this situation, the sum of the frequencies of the signal and idler
photons is fixed, thus the photons' frequencies are
anticorrelated.  SPDC with generalized joint spectral properties
({\it i.e.} correlated, uncorrelated, anticorrelated) was studied
theoretically by Campos {\it et al.}~\cite{Campos90}.  Here we
review the experimental proposals for generating
frequency-uncorrelated SPDC.

A number of techniques have been proposed for creating
frequency-uncorrelated SPDC; however, they all impose certain
constraints on the center frequencies and/or bandwidths of the
SPDC photons. Grice {\it et al.} proposed a method for creating
frequency-uncorrelated SPDC based on a group-velocity matching
condition introduced by Keller and Rubin~\cite{Keller97}. Their
method can be used to create degenerate, frequency-uncorrelated
photons; however, the center frequency of down-conversion is fixed
by the nonlinear material, and the bandwidth of two SPDC photons
must be equal. They also demonstrate that degenerate,
frequency-uncorrelated photons with different bandwidths may be
generated; however, in this case the bandwidths are fixed and
cannot be independently controlled.
 Giovannetti {\it et al.} proposed extending the approach of Grice
{\it el al.} by using a periodically poled nonlinear
crystal~\cite{Giovannetti02b}.  This allows one to to satisfy the
zeroth-order term in the phase-matching relation at an arbitrary
pump wavelength, making the group-velocity matching relation
easier to satisfy. Even with such an enhancement, this approach
does not have sufficient flexibility to allow independent control
of the marginal spectra.

A distinct approach for creating frequency-uncorrelated SPDC was
independently discovered by U'Ren {\it et al.}~\cite{Uren03} and
our group~\cite{Walton03c}. Instead of relying on the satisfaction
of a group-velocity matching condition, these approaches rely on
the geometrical symmetry of degenerate, non-collinear Type-I SPDC
(the previously mentioned techniques only worked with Type-II
SPDC). The essential difference between the two proposals is that
for U'Ren {\it et al.}, the phase-matching relation in the pump
propagation direction is a constraint that must be satisfied,
while for our auto-phase-matched technique, the single-mode
waveguide ensures that this relation is satisfied, regardless of
the system parameters.  The relative lack of constraints for these
techniques makes them attractive, since it suggests that they
remain viable options even if the center frequency of SPDC is
constrained by some other factor (optical fiber loss, detector
efficiency, etc.). Nonetheless, both techniques suffer a lack of
flexibility that is reminiscent of the previously mentioned
schemes.  The SPDC photons must be degenerate, and the bandwidths
of the two photons must be equal. In the next section, we show
that by generalizing our auto-phase-matched technique, we can
obtain independent spectral control of the SPDC photons.

To begin, we review the relationship between the pump pulse and
the SPDC in the geometry of Fig.~\ref{setup}.  Following the
derivation in Ref.~\cite{Booth02}, a classical pump pulse
described on the free-space side of the waveguide-air interface by
\begin{equation}\label{pump}
E_p(z,t)\propto \int\!\!\!\int dk\, d\omega\,
\tilde{E_p}(k,\omega)e^{-i(kz-\omega t)}
\end{equation}
stimulates the creation of a pair of photons described by the
two-photon wavefunction
\begin{equation}
|\Psi\rangle \propto \int\!\!\!\int d\omega_{s}\, d\omega_i\,
\phi\left(\omega_i,\omega_s\right)\,|\omega_s\rangle_s\,|\omega_i\rangle_i,
\end{equation}
where
\begin{equation}\label{wavefunction}\phi\left(\omega_i,\omega_s\right)=\tilde{E_p}\left[\frac{\beta_i(\omega_i)-\beta_s(\omega_s)}{n_p(\omega_i+\omega_s)},\omega_i+\omega_s\right]
\end{equation}
$n_p(\omega_i+\omega_s)$ is the refractive index for the pump
polarization.  Here, and for the rest of the paper, we use the
variable $k$ to refer to the component of the pump wavevector
along the $z$-axis.  The ket
$|\omega_s\rangle_s\,|\omega_i\rangle_i$ represents a signal
photon in the frequency mode $\omega_s$ and an idler photon in the
frequency mode $\omega_i$ with corresponding propagation constants
$\beta_s(\omega_s)$ and $\beta_i(\omega_i)$, respectively.
Equation~(\ref{wavefunction}) conveys the main result of this
paper:  Assuming the dispersion properties of the medium are
known, it is possible to generate a down-converted photon pair
with arbitrary joint spectrum by appropriately engineering the
spatial and temporal characteristics of the pump pulse.

In Fig.~\ref{setup}A, the pump pulse is parameterized by three
numbers: the center frequency, the temporal coherence length
$\sigma_x$, and the spatial coherence length $\sigma_z$. These
parameters may be chosen to produce degenerate SPDC with
controllable entanglement, as described in Ref.~\cite{Walton03c};
however, in order to obtain independent control of the center
frequency and bandwidth of each SPDC photon, one must relax the
constraints on the pump pulse, as in Fig.~\ref{setup}B. Using
Eqs.~(\ref{pump}) and (\ref{wavefunction}), it is straightforward
to show that a pump pulse described by
\begin{widetext}
\begin{equation}\label{genpump}
\tilde{E_p}(k,\omega) \propto
\mbox{Exp}\left[-\left(\frac{\left(n_pk-k_p\right)+\left(\omega-\omega_p\right)\beta_s^{'}}{2\beta^{'}\sigma_i}\right)^2-\left(\frac{\left(n_pk-k_p\right)-\left(\omega-\omega_p\right)\beta_i^{'}}{2\beta^{'}\sigma_s}\right)^2\right]
 \end{equation}
will yield the following frequency-uncorrelated two-photon state
\begin{equation}\label{genwavefunction}
|\Psi\rangle \propto \int\!\!\!\int d\omega_{s}\, d\omega_i\,
\mbox{Exp}\left[-\left(\frac{\omega_i-\omega^{(\circ)}_i}{2\sigma_i}\right)^2-\left(\frac{\omega_s-\omega^{(\circ)}_s}{2\sigma_s}\right)^2\right]\,|\omega_s\rangle_s\,|\omega_i\rangle_i,
\end{equation}
\end{widetext}
where $\omega^{(\circ)}_s$ and $\omega^{(\circ)}_i$ are the center
frequencies of the signal and idler beams, respectively, and we
have used the following definitions and approximations:
\begin{eqnarray}
k_p&\equiv&\beta_i(\omega^{(\circ)}_i)-\beta_s(\omega^{(\circ)}_s)\\
\omega_p&\equiv&\omega^{(\circ)}_s+\omega^{(\circ)}_i\\
\beta_j(\omega_j)&\approx&\beta_j(\omega^{(\circ)}_j)+(\omega_j-\omega^{(\circ)}_j)\beta^\prime_j\quad j=s,i\\
n_p(\omega_i+\omega_s)&\approx&n_p(\omega_p)\equiv n_p\label{end}.
\end{eqnarray}
These approximations are valid in typical situations; however, if
required, more terms may be used at the expense of a more
complicated expression for the pump pulse.

Equations~(\ref{genpump}) and (\ref{genwavefunction}) summarize
the central result of this work.  Taken together, these relations
can be thought of as an algorithm for producing
frequency-uncorrelated SPDC with arbitrary marginal spectra.  The
wavefunction in Eq.~(\ref{genwavefunction}) describes a
frequency-uncorrelated two-photon state in which the signal photon
is centered on $\omega^{(\circ)}_s$ with a bandwidth $\sigma_s$,
and the idler photon is centered on $\omega^{(\circ)}_i$ with a
bandwidth $\sigma_i$.  Note that these two photons are not
themselves indistinguishable (unless
$\omega^{(\circ)}_s=\omega^{(\circ)}_i$ and $\sigma_s=\sigma_i$).
As previously mentioned, the indistinguishability arises in a
multi-photon experiment when one photon of the pair enters an
interferometer with one or more photons that have identical
spectra.  In this case, the lack of frequency correlations between
the signal and idler prevents a loss of interferometric visibility
by ensuring that spectral measurements on one photon of the pair
won't reveal any spectral information about the other.

Equations~(\ref{genpump}-\ref{end}) demonstrate that the four
numbers $\omega^{(\circ)}_s$, $\omega^{(\circ)}_i$, $\sigma_s$,
and $\sigma_i$, along with the dispersion properties of the
waveguide, are sufficient to determine the form of the pump pulse
required to generate the desired wavefunction.  We can simplify
the description of the pump pulse by rewriting Eq.~(\ref{genpump})
as
\begin{equation}\label{genpump2}
\tilde{E_p}(k,\omega)
=\mbox{Exp}\left[-\left(\frac{\omega-\omega_p}{2A}\right)^2-\left(\frac{\left(k-k_p/n_p\right)+C\left(\omega-\omega_p\right)}{2B}\right)^2\right],
\end{equation}
where
 \begin{eqnarray}\label{genpump5}
A&=&\frac{\beta^\prime_s+\beta^\prime_i}{\sqrt{\left(\frac{\beta^\prime_s}{\sigma_i}\right)^2+\left(\frac{\beta^\prime_i}{\sigma_s}\right)^2-\frac{\left(\beta^\prime_s\sigma_s^2-\beta^\prime_i\sigma_i^2\right)^2}{\left(\sigma_s\sigma_i\right)^2\left(\sigma_s^2+\sigma_i^2\right)}}}\\
B&=&\frac{\beta^\prime_s+\beta^\prime_i}{n_p\sqrt{\frac{1}{\sigma_s^2}+\frac{1}{\sigma_i^2}}}\\
C&=&\frac{\beta^\prime_s\sigma_s^2-\beta^\prime_i\sigma_i^2}{n_p\left(\sigma_s^2+\sigma_i^2\right)}.
\end{eqnarray}
The algorithm for creating the appropriate pump pulse to produce
the state in Eq.~(\ref{genwavefunction}) is as follows. A pulse is
created with center frequency $\omega_p$, spectral bandwidth $A$
and spatial bandwidth $B$.  Next, a dispersive element such as a
wedge of quartz or a diffraction grating is used to correlate $k$
and $\omega$ by effecting the substitution
\begin{equation}
k\rightarrow k+C(\omega-\omega_p).
\end{equation}
Finally, the pulse it directed towards the nonlinear waveguide at
incidence angle
\begin{equation}\label{theta}
\theta=\sin^{-1}\frac{k_pc}{n_p\omega_p},
\end{equation}
where $\theta$ is measured outside the waveguide (see
Fig.~\ref{setup}B), and $c$ is the speed of light in vacuum.

\begin{figure*}[t]
\includegraphics{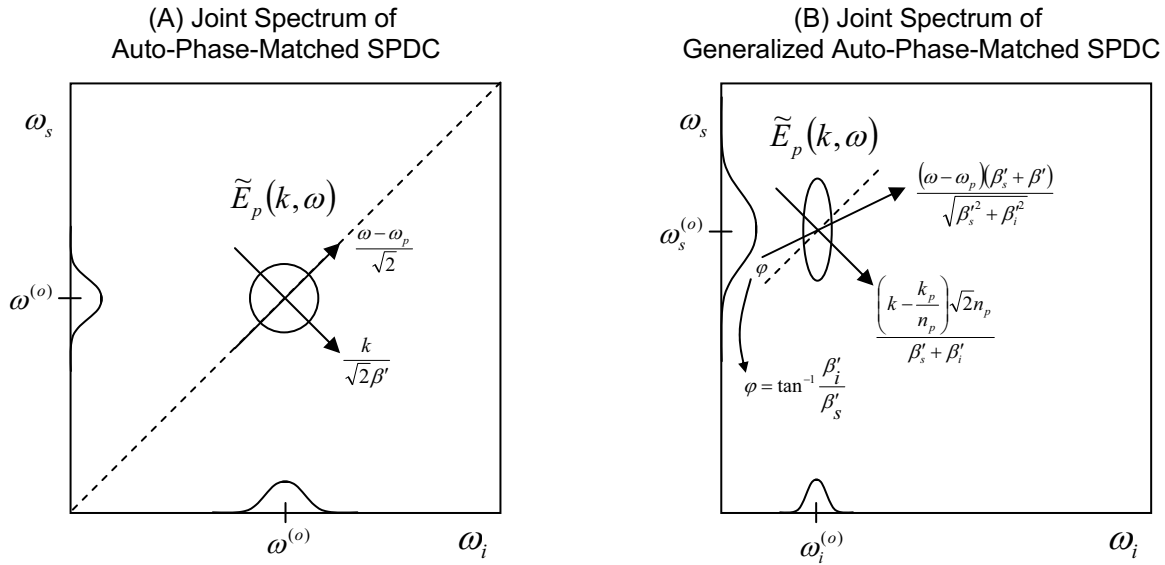}
\caption{A graphical technique for associating the
spectral/spatial properties of a pump pulse
($\tilde{E_p}(k,\omega)$, plotted on the inner axes), with the
joint spectrum of the resulting down-converted photon pair (the
outer axes), for auto-phase-matched SPDC (A) and generalized
auto-phase-matched SPDC (B).  In both (A) and (B),
$\tilde{E_p}(k,\omega)$ is plotted on axes that are both rotated
and scaled by constants related to the waveguide dispersion
properties, such that the resulting plot can be interpreted using
the outer axes. The central point of this paper is conveyed in
(B). That is, by judicious choice of $\tilde{E_p}(k,\omega)$, one
can generated non-degenerate, frequency-uncorrelated SPDC, such
that the bandwidth of each photon is independently controllable.
The marginal spectrum of each photon is plotted along the
appropriate axis in each figure to demonstrate the connection
between cross-spectral correlation in the pump and independent
control of the photons' bandwidths.}\label{jointspectrum}
\end{figure*}

In Fig.~\ref{jointspectrum}, we present a graphical depiction of
the relationship between the pump pulse and the resulting
two-photon state, for both auto-phase-matched SPDC
(Fig.~\ref{jointspectrum}A) and generalized auto-phase-matched
SPDC (Fig.~\ref{jointspectrum}B).  In both cases, a plot of
$\tilde{E_p}(k,\omega)$ is superposed over the joint spectrum of
the signal and idler photons.  By plotting $\tilde{E_p}(k,\omega)$
at the correct location and on the correct inner axes, one can
immediately infer the joint spectrum of the down-converted
photons, simply by interpreting the plot using the outer axes. In
Fig.~\ref{jointspectrum}A, the non-zero portion of
$\tilde{E_p}(k,\omega)$ is centered on the $\omega_s=\omega_i$
axis, and one of the inner axes is scaled by the factor
$\beta^\prime$, which is the first derivative of $\beta(\omega)$
evaluated at $\omega^{(\circ)}$.  In Fig.~\ref{jointspectrum}B,
the non-zero portion of $\tilde{E_p}(k,\omega)$ is located at a
general position, and the inner axes are no longer orthogonal
(unless $\beta_s^\prime=\beta_i^\prime$).  Using this figure, the
two desirable features of the two-photon joint spectrum
(non-degeneracy and independently controlled bandwidths) are
easily interpreted in terms of the pump pulse.  That is, the
non-degeneracy of the photon pair derives from the condition
$k_p\neq 0$, which in turn derives from the non-normal incidence
of the pump pulse.  Similarly, the independent control of the two
photons respective bandwidths derives from the cross-spectral
correlation in the pump pulse ($\tilde{E_p}(k,\omega)$ does not
factor into a function of $k$ times a function of $\omega$).

\section{EXAMPLE: POLARIZATION-ENTANGLED FREQUENCY-UNCORRELATED
SPDC FROM A BBO WAVEGUIDE}\label{example}

\begin{table}\label{table}
\begin{tabular}{p{1.5in}p{.75in}p{.75in}}\hline\hline
&$\tilde{E_p^z}(k,\omega)$&$\tilde{E_p^y}(k,\omega)$\\\hline
$A$ ($10^{12}$ rad/s)&1.89&1.89\\
$B$ ($10^3$ rad/m)&1.35&1.25\\
$C$ ($10^{-9}$ s/m)&3.54&3.28\\
$\theta$ ($^\circ$)&$-20.1$&$-18.6$\\\hline\hline
\end{tabular}\caption{Parameters that describe the pump pulse required to produce non-degenerate, frequency-uncorrelated,
polarization-entangled SPDC in a single-mode BBO waveguide, using
the technique depicted in Fig.~\protect\ref{setup}B.  The signal
photon is at $0.8$ $\mu$ with coherence length $1$ mm, and the
idler photon is at $1.5$ $\mu$ with coherence length $1$ cm. The
pump pulse is comprised of coherently superposed, independently
controlled pulses $\tilde{E_p^z}(k,\omega)$ and
$\tilde{E_p^y}(k,\omega)$ in the two polarization modes $z$ and
$y$, respectively. The parameters $A$, $B$, $C$, and $\theta$ are
defined in Sec.~\protect\ref{gapmspdc}.  The negative values of
$\theta$ indicate that the projection of the pump wavevector along
the waveguide is oriented in the negative $z$ direction (see
Fig.~\protect\ref{setup}B).}
\end{table}

In the generalized auto-phase-matched technique, one can obtain
polarization entanglement by adjusting the polarization state of
the pump pulse, without sacrificing the spectral control described
above. To illustrate this feature, we present an example involving
non-degenerate, polarization-entangled SPDC produced in a
single-mode BBO waveguide.

The general idea is to use two of the nonlinear medium's
$\chi^{(2)}$ tensor elements at the same time, by preparing a
coherent superposition of two polarization modes of the pump
pulse. When producing polarization-entangled photon pairs, it is
typically desirable that a given photon have the same spectral
properties for both two-photon polarization amplitudes. Therefore,
in creating the pump pulse, we use the same four numbers
$\omega^{(\circ)}_s$, $\omega^{(\circ)}_i$, $\sigma_s$, and
$\sigma_i$ in calculating the desired pulse shape for both pump
polarization modes. However, since the two two-photon amplitudes
relate to SPDC processes taking place in distinct polarization
modes, the dispersion properties of the waveguide will in general
be different.  Thus, using the notation of Fig.~\ref{setup}, the
pump pulse will be characterized by two functions:
$\tilde{E_p^y}(k,\omega)$, which describes the $y$-polarized
component of the pump-pulse, and $\tilde{E_p^z}(k,\omega)$, which
describes the $z$-polarized component of the pump-pulse

In the case of BBO, the relevant tensor elements are
$\chi^{(2)}_{yyy}=2.22$ pm/V and $\chi^{(2)}_{zxx}=0.16$
pm/V~\footnote{Since $\chi^{(2)}_{yyy}$ and $\chi^{(2)}_{zxx}$ are
not equal, we adjust the ratio of the optical powers in each pump
polarization mode in order to obtain the maximally entangled state
$|\mbox{HH}\rangle +e^{i\phi}|\mbox{VV}\rangle$, where the
relative phase $\phi$ is determined by the relative phase between
the two pump polarization modes. It is straightforward to produce
non-maximally entangled and/or mixed polarization states with this
technique. Non-maximally entangled states may be produced by
adjusting the ratio of the optical powers in each of the pump's
polarization modes.  Mixed states may be produced by allowing the
pump pulse to become partially depolarized.  Although in this
paper we are concerned with frequency-uncorrelated (and thus,
frequency-unentangled) photon pairs, analogous generalizations for
partial entanglement and mixedness in the frequency degrees of
freedom are possible, given the appropriate manipulations of the
pump pulse.}. Therefore, using the notation of Fig.~\ref{setup},
the pump beam will approach the waveguide in the $x$-$z$ plane,
and will be composed of a $y$-polarized pulse and a $z$-polarized
pulse.  In Table~I, we list the calculated values of $A$, $B$,
$C$, and $\theta$ (defined in Sec.~\ref{gapmspdc}) that correspond
to a frequency-uncorrelated polarization-entangled pair of photons
with the signal photon at $0.8$ $\mu$ with coherence length $1$
mm, and the idler photon at $1.5$ $\mu$ with coherence length $1$
cm. These values for center wavelength and coherence length were
chosen in order to make the signal photon suitable for
long-distance optical fiber transmission, and the idler photon
suitable for local processing in an integrated optical circuit. In
calculating the values in Table~I, we have ignored waveguide
dispersion, using instead the Selmeier curves to describe the
material dispersion in the BBO single-mode waveguide.

\begin{figure}[t!]
\includegraphics[scale=.45]{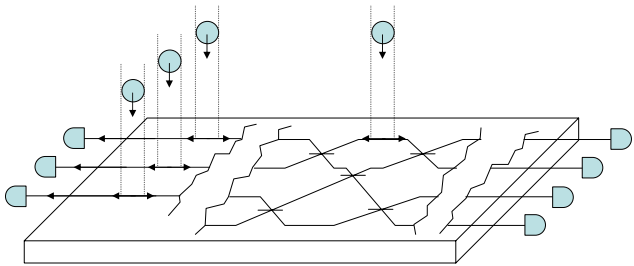}
\caption{A conceptual schematic of an integrated optical quantum
information processor.  The entire integrated circuit is
constructed on a nonlinear material such that any strech of
waveguide may be used as a source of counter-propagating photons
pairs.  In the text, we describe how generalized
auto-phase-matched SPDC is particularly well-suited for this
application.}\label{integrated}
\end{figure}

\section{QUANTUM INFORMATION PROCESSING WITH INTEGRATED OPTICAL
CIRCUITS}\label{int}

Generalized auto-phase-matched SPDC is particularly well-suited
for quantum information processing on an integrated optical
circuit (see Fig.~\ref{integrated}).  Among the advantages of
replacing an array of discrete optical elements with an integrated
optical circuit are the following: reduced size, reduced loss due
to fewer connectors, and ``common mode'' noise processes because
of the close proximity of optical elements.  However, there are
substantial experimental challenges associated with constructing
an integrated optical quantum information processor.  Perhaps the
most obvious challenge is finding a material that can perform as
many of the required functions (photon source, modulation,
detection) as possible.  A significant advantage of generalized
auto-phase-matched SPDC in that the choice of material places
essentially no limitation on the spectral and polarization
properties of the photon pairs that will be produced.  All that is
required is that the material's $\chi^{(2)}$ tensor have the
appropriate non-zero elements such that, for a given orientation
of the optic axis with respect to the waveguide, the desired SPDC
process will occur.

Figure~\ref{integrated} depicts a conceptual schematic of an
integrated optical quantum information processor which employs
generalized auto-phase-matched SPDC for generating photons.  The
figure highlights several of the practical advantages associated
with this technology.  First, the sources may be placed at the
edge of the circuit and combined with single-photon counters to
implement conditional single-photon sources. Second, due to the
transverse-pump configuration, the photon pairs may be created
within the interior of the optical circuit.  Finally, since there
is no group-velocity matching relation associated with generalized
auto-phase-matched SPDC, poling of the nonlinear waveguide at each
source is not required.

\section{CONCLUSIONS}\label{conclusions}

We have described a scheme for generating polarization-entangled
photons pairs with arbitrary spectrum. By controlling the spatial,
temporal, and polarization properties of the pump pulse, it is
possible to generate the desired two-photon state, regardless of
the dispersion properties of the nonlinear medium. We provided a
calculation of the parameters describing the pump pulse required
to generate a photon-pair with a particular joint spectrum in a
single-mode BBO waveguide. Finally, we discussed the role this
source technology might play in a distributed quantum information
processor based on integrated optics.

\section*{ACKNOWLEDGMENTS}

 This work was supported by the National Science Foundation;
the Center for Subsurface Sensing and Imaging Systems (CenSSIS),
an NSF Engineering Research Center; the Defense Advanced Research
Projects Agency (DARPA); and the David and Lucile Packard
Foundation.


\end{document}